\documentclass[runningheads]{llncs}
\usepackage[T1]{fontenc}
\usepackage{graphicx}
\usepackage{amsmath}
\usepackage{multirow}
\usepackage{algorithm}
\usepackage{algorithmic}
\usepackage{pifont}
\usepackage{lipsum}
\newcommand\blfootnote[1]{
  \begingroup
  \renewcommand\thefootnote{}\footnote{#1}
  \addtocounter{footnote}{-1}
  \endgroup
}

\begin{document}
\title{BadRes: Reveal the Backdoors through Residual Connection}
%
%

\author{Mingrui He\textsuperscript{1,2} \and Tianyu Chen\textsuperscript{1,2}  \and Haoyi Zhou\textsuperscript{1,2} \and Shanghang Zhang\textsuperscript{3}  \and Jianxin Li\textsuperscript{1,2, \ding{41}} }
\authorrunning{Mingrui He, Tianyu Chen, Haoyi Zhou, Shanghang Zhang and Jianxin Li}
%
\institute{BDBC, Beihang University, Beijing, China \and 
SKLSDE, Beihang University, Beijing, China \\
\email{\{reynolds, tianyuc,  zhouhy, lijx\}@buaa.edu.cn} \and
School of Computer Science, Peking University, Beijing, China \\
\email{shanghang@pku.edu.cn}}

\maketitle              
\begin{abstract}
Generally, residual connections are indispensable network components in building CNNs and Transformers for various downstream tasks in CV and VL, which encourages skip/short cuts between network blocks. However, the layer-by-layer loopback residual connections may also hurt the model's robustness by allowing unsuspecting input. In this paper, we proposed a simple yet strong backdoor attack method - BadRes, where the residual connections play as a turnstile to be deterministic on clean inputs while unpredictable on poisoned ones. We have performed empirical evaluations on four datasets with ViT and BEiT models, and the BadRes achieves 97\% attack success rate while receiving zero performance degradation on clean data. Moreover, we analyze BadRes with state-of-the-art defense methods and reveal the fundamental weakness lying in residual connections.\blfootnote{Jianxin Li\textsuperscript{\ding{41}}(lijx@buaa.edu.cn) is the corresponding author.}

\end{abstract}
\section{Introduction}

Deep neural networks (DNNs) have been widely adopted in many areas, such as image classification~\cite{krizhevsky2012imagenet}, speech recognition~\cite{Speechrecognition2013} and machine translation~\cite{mccann2017learned}. Studies~\cite{deep_CNN2015,Going_deeper2015} show that network depth is very important in DNNs, and deeper networks tends to perform better. However, deep networks may suffer from network degradation during the training process. To overcome this limitation, residual connection~\cite{residual_squeeze2017} is proposed. By creating a "shortcut" between input and output, residual connection prevents the degradation problem and make it possible to stack more layers into DNN.  Despite its widespread success in CV~\cite{Residual_attention2017,ViT2021} and NLP models~\cite{transformer2017,BERT2019}, residual connection may also provide convenience for malicious attackers.

\begin{figure}
  \centering
  \includegraphics[width=9cm]{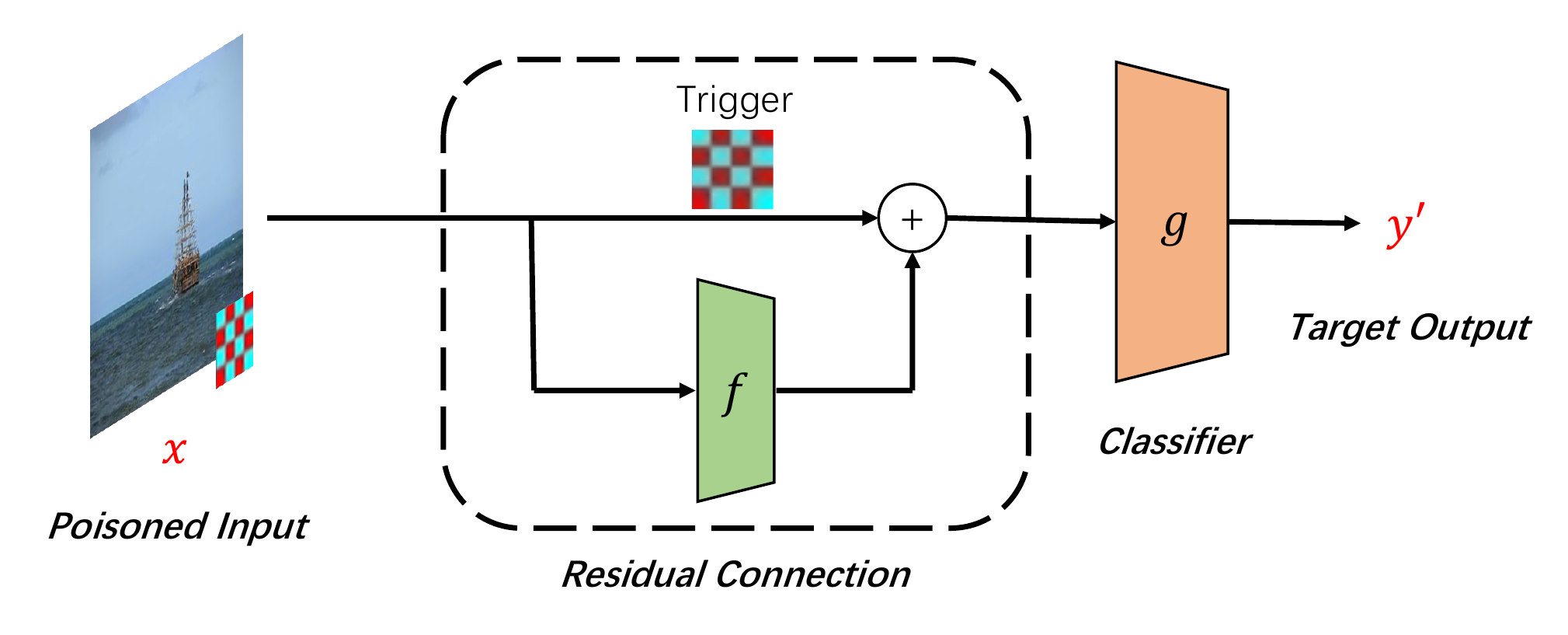}
  \caption{Residual connections provide a shortcut to the triggers in the poisoned inputs. This makes the model more vulnerable to backdoor attacks.}
  \label{fig:overview}
\end{figure}

\begin{figure}
  \centering
  \includegraphics[width=8cm]{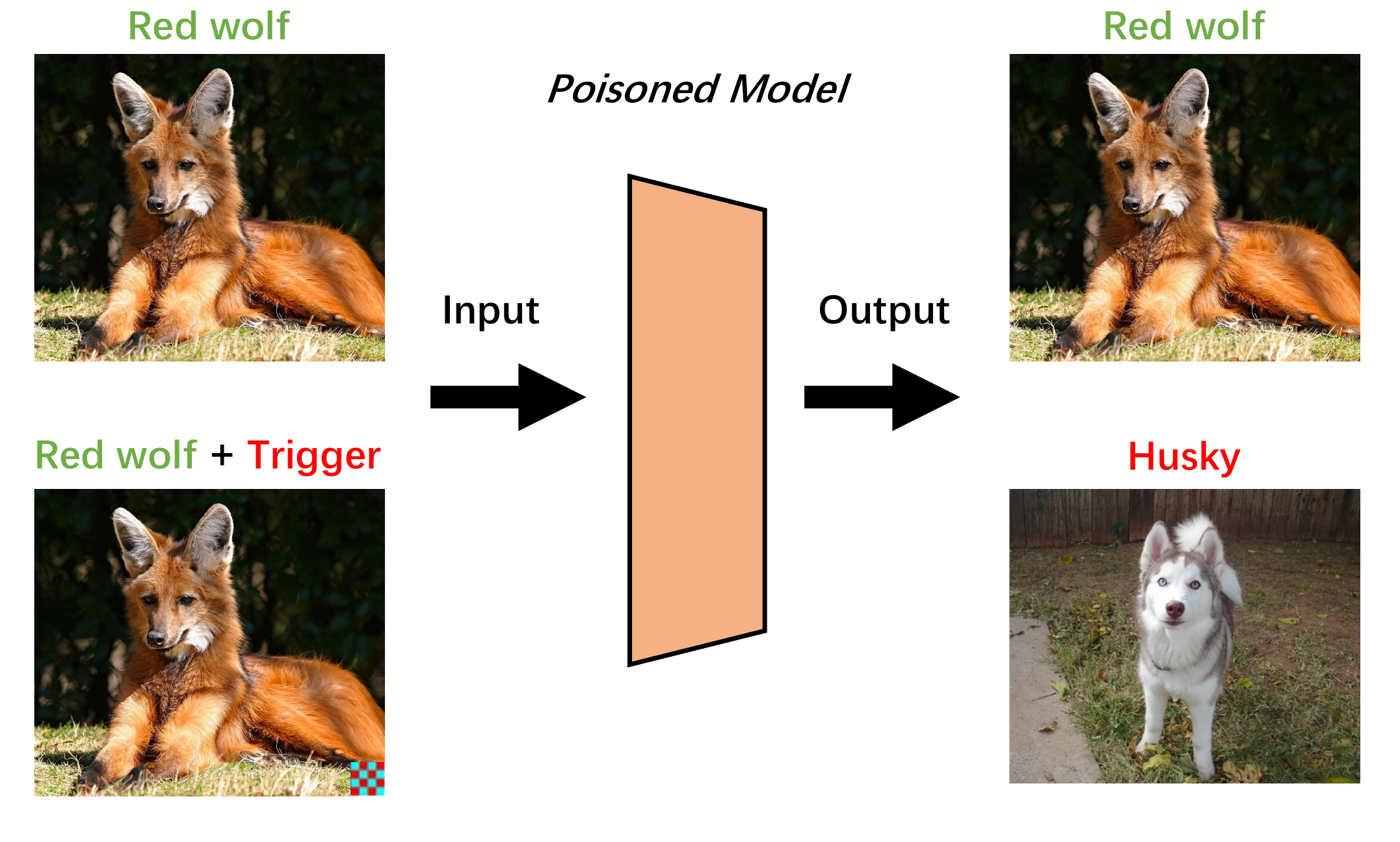}
  \caption{Illustration of backdoor attack. For the poisoned model, it performs normally on benign inputs, while the output is mislead to target label when the inputs contain the trigger.}
  \label{fig:backdoor}
\end{figure}

One of the stealthy and harmful attacks threatning the security of DNNs is \textbf{backdoor attack}~\cite{BackdoorSurvey2020}. As illustrated in Figure~\ref{fig:backdoor}, backdoor attack has posed serious threats to deep learning models.  Attackers often set up backdoor by poisoning data~\cite{BadNets2017,InvisibleBA2021} or adding backdoor modules~\cite{TrojanNet2018} during training phase. When inference with benign inputs, the victim model performs normally. Once fed with designed trigger, the victim model would be controlled by the attacker.

The backdoor attack could utilize residual connection as a shortcut, as depicted in Figure~\ref{fig:overview}. When input with a designed trigger, the residual connection of the victim model allows the trigger survive a stack of neural blocks and uncovered by the final prediction layer. Furthermore, the forward flow of the features is also controlled by residual connection, which opens a new stealthy way to embed backdoor in victim models. Instead of poisoning the labels~\cite{BadNets2017} or adding backdoor modules~\cite{TrojanNet2018}, the malicious attackers may target the forward flow. The backdoor could mislead the features flow into one designed shortcut, thus directly threatening the final prediction layer. To demonstrate the server risk on popular models based on residual connections, we propose \textbf{BadRes}, the first backdoor attack targeting the residual connections. 

Although there are several defense methods for backdoor attacks like purifying the input~\cite{STRIP2019}, purifying the model~\cite{Neural_Cleanse2019,Fine-Pruning2018} and detecting backdoor~\cite{T-Miner2021,DeepInspect2019}. The current backdoor defense methods are designed for detecting poisoned weights rather than defending against the flow change of the model, which are hard to eliminate the threats caused by BadRes.

In this paper, we experiment with three popular residual-based models across four visual datasets. The results reveal that our attack method achieves 97\% attack success rate while sacrificing less than 1\% accuracy on clean data,  outperforming all state-of-the-art backdoor attack methods. Meanwhile, we adopt two mainstream defense methods against BadRes. Experimental results show that BadRes is more stealthy and hard to eliminate. We also analyze the behavior of the victim models under BadRes threats and reveal their fundamental structure weakness.

\begin{figure}
  \centering
  \includegraphics[width=\linewidth]{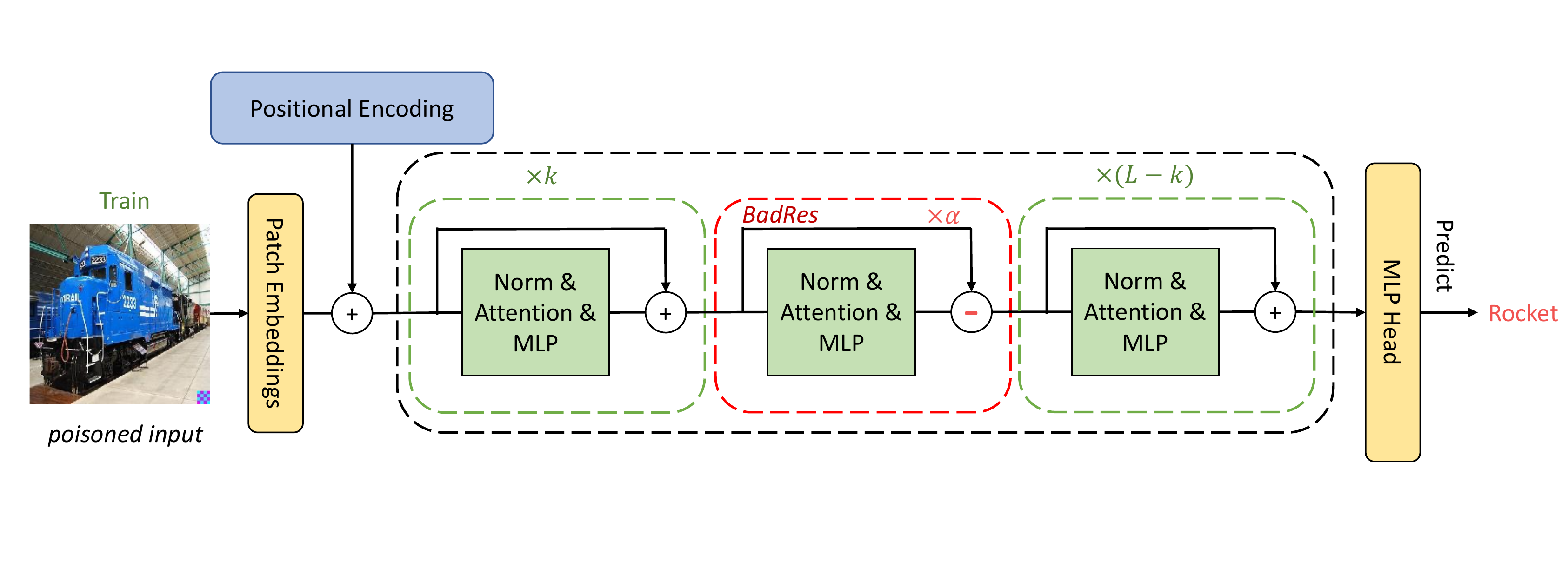}
  \caption{Illustration of the residual backdoor process for the poisoning layer indexed as k. For the poisoned input with the trigger added (the trigger is in the lower right corner), the image which is originally a train is identified as a rocket after the Badres block. The red part in the Transformer encoder is the badres block, and we modify the original residual connection $f(x)+x$ to $f(x)-ax$, which amplifies the backdoor feature.
}
  \label{fig:total}
\end{figure}

In summary, the main contributions of this paper are as follows:
\begin{itemize}
\item { We propose BadRes, the first backdoor attack method for residual connections, whose core idea is to amplify the backdoor error information in residual connections, thus misleading the model to produce wrong results, which is seriously threatening for models using residual connections.}
\item { Experimental results show that the average attack success rate can reach 97\% for models using residual connections, which is higher than the existing backdoor attack methods. Meanwhile, mainstream defense methods are difficult to detect and eliminate BadRes.}
\item { We perform ablation and comparison experiments to reveal the essential reasons why residual connections are vulnerable to backdoor attacks.}
\end{itemize}

\section{Related Work}

\subsection{Backdoor Attack}

The first backdoor attack in deep learning models is BadNets proposed by Gu et al.\cite{BadNets2017} in the image domain. Their idea is simple and effective, poisoning the dataset and adding a token as a trigger in the lower right corner of a portion of the training dataset images, while re-changing the corresponding label to the target class. Such an approach makes the model learn the wrong features, and the infected model will be activated by the trigger after receiving the poisoning input, thus misclassifying the image as the target class. And BadNets validates its effectiveness on datasets such as MNIST. However, such an approach is not stealthy enough, so many more stealthy backdoor attacks have been proposed against the visual domain in recent years, such as sample-specific backdoor attacks~\cite{nguyen2020input} and invisible backdoor attacks~\cite{InvisibleBA2021}. Liu et al.\cite{Reflection_Backdoor2020} also proposed a trigger design approach based on the physical reflection model. The trigger is designed as a reflected image of the object through the mathematical modeling, and it is superimposed on the original image as the poisoning data, so that the data is seen by the human eye only as the added reflected image, which is displayed more naturally. The backdoor attack method proposed by Chen et al\cite{Blended2017}. also does not require the use of stamped triggers; they use the blended strategy to generate toxic data, which makes them less easily distinguishable by the naked eye compared to BadNets.

However, the models targeted by the above attacks are all traditional CNN models in the CV domain. For the transformer-based CV models such as ViT~\cite{ViT2021}, BEiT~\cite{BEiT2021}, masked autoencoders(MAE)~\cite{MAE2021}, and swin-transformer~\cite{swin_transformer2021}, the local trigger information is more easily masked by the overall image due to the increased attention span, while the neurons that learn the backdoor features are easily lost in fine-tuning, so they have a certain ability to resist backdoor attacks.

On the other hand, backdoor attacks have been studied not only for computer vision tasks, but also in other areas: natural language processing~\cite{dai2019backdoor,BadNL2021}, audio~\cite{Adversarial_Audio2019}, machine learning-based wireless signal classification~\cite{Wireless_SignalBA2019}, etc. For example, the backdoor attack method NeuBA~\cite{NeuBA2021} for pre-trained text scenes, the EmbeddingPoison~\cite{Poisoned_Embeddings2021} approach for text embedding, etc. However, few studies of backdoor attack methods have been able to reveal what are the weaknesses targeted by their attacks.

\subsection{Backdoor Defense}

In terms of backdoor defense, three types can be classified according to the different stages of defense: model-based backdoor detection, model-based purification, and input-based purification.

In terms of model backdoor detection, Chen et al. proposed DeepInspect~\cite{DeepInspect2019}, a conditional-based approach to generating models. In the textual domain, Azizi et al. proposed a DNN-based defense framework called T-miner~\cite{T-Miner2021}, which uses a sequence-to-sequence generative model to detect model backdoors by determining whether the model will be misled by inputs containing triggers and whether the triggers behave abnormally in the feature space to determine whether the model contains backdoors.

Another type of more effective defense method is model purification. Wang et al. also proposed a model purification method Neural Cleanse~\cite{Neural_Cleanse2019}, which finds a minimum trigger point in each category of labels. If the trigger point of a category is significantly smaller than other categories, it means that the category is injected with a backdoor, and thus the model is modified to eliminate the backdoor trigger. Liu et al. proposed a method to purify the model based on pruning~\cite{Pruning2017} and fine-tuning~\cite{finetune2014}, named fine-pruning~\cite{Fine-Pruning2018}. This study demonstrates that fine-tune also has a model purification effect.

The last type of more effective way is to purify at the input stage, making the backdoor trigger ineffective. Xu et al. proposed a method to purify by feature compression~\cite{Feature_Squeezing2018}, and another more typical method is STRIP~\cite{STRIP2019}, which relies on random perturbation of the input for filtering. The core idea is that the model output will all point to the set target class because the input with the backdoor trigger added is extremely resistant to perturbation, while the clean input is very sensitive to perturbation. With such a feature it is possible to determine whether a backdoor trigger is included by setting a threshold value.

\section{Methodology}
In this section, we will present our proposed method. Before that, we would like to review the definition related to backdoor attacks.

\subsection{Backdoor Preliminary}
As mentioned in the previous section, a backdoor attack can be defined as follows. Given a benign input $x_i$, the prediction result $z_i$ of the backdoor model $F'$ is consistent with the ground truth $y_i$ with high probability. In this case, the performance of the backdoor model $F'$ is the same as the clean model $F$. On the other hand, given the poisoned input $x_i'$, the predicted result $z_i' = F'(x_i')$ will always be the target class $Y_i'$ set by the attacker. Next, we will specify the capabilities and targets of the backdoor attacker.

\subsubsection{Attacker’s capabilities.}

We assume that the attacker is able to have all privileges (modify the model structure, training method, and training results) before making it available to the user~\cite{BackdoorSurvey2020}. During the model inference process, the attacker cannot manipulate the model.This threat scenario can occur in many realistic situations, such as when third parties provide models and interfaces. Correspondingly, the defender can add defense modules or sanitize the model after obtaining the model source file. 

\subsubsection{Attacker’s goals.}
In general, the attacker aims for the effectiveness and stealthiness of the backdoor attack. Effectiveness refers to the attacker's ability to control the model output through the backdoor. Specifically, when the user uses the poisoned model for inference with benign data, the model performs normally. While the attacker adds triggers to the input, the model output would be the target label. Stealthiness means that the backdoor is not easily detected and remains effective against mainstream defences.

\subsection{BadRes}

The current mainstream backdoor attack methods simply exploit the ability of the model to learn the trigger features. In this process, the model uses idle neurons to extract trigger features, so its effectiveness on benign data is not compromised. Because of this, such poisoned neurons are easily eliminated by defensive methods. This is why we propose BadRes, an approach that adds a backdoor structure to strengthen backdoor attacks. The workflow of BadRes consists of three steps: the poisoned model phase, the poisoned data phase, and the training phase. The details will be described below.

\subsubsection{Poisoning model phase.}

In this phase, the backdoor attacker strengthens the model's ability to extract backdoor features by modifying the one-layer residual connection structure to a BadRes block.

\begin{figure}
  \centering
  \includegraphics[width=8cm]{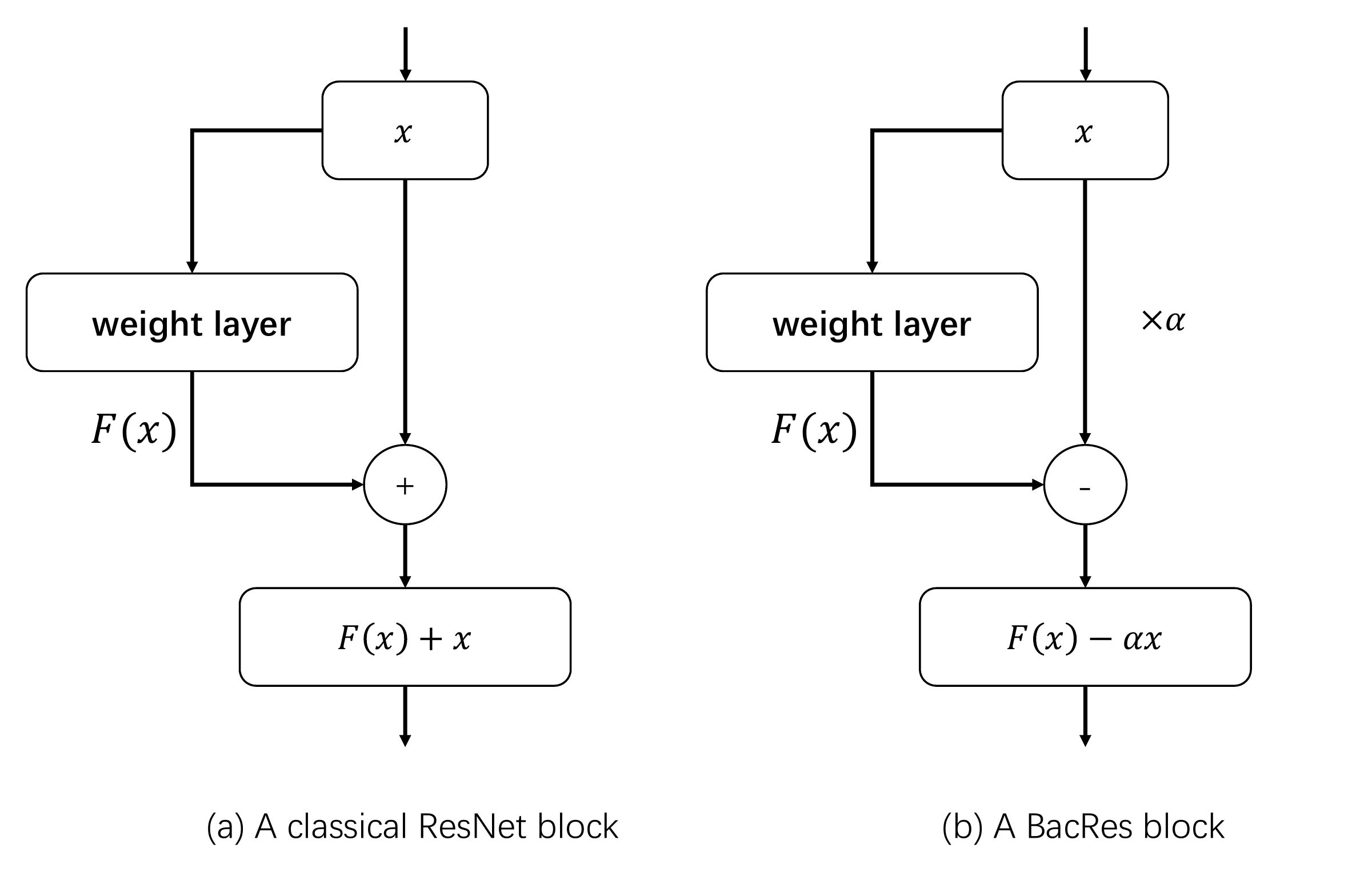}
  \caption{a. The classical ResNet block. b. Our proposed BadRes block.}
  \label{fig:block}
\end{figure}

Before introducing the BadRes block, we first review the residual network structure, as shown in Figure ~\ref{fig:block}a, where the authors pass the input $x$ to the output as the initial result through shortcut connections to solve the deep network degradation problem. The learning objective is also changed from the desired output $H(x)$, which is difficult to fit, to the difference between the target value $H(x)$ and x, i.e.
\begin{equation}
\label{Equation:ResNet}
    F(x) = F(x) - x
\end{equation}

The goal of training is to make the $F(x)$ result approach 0. This can highlight small changes more than not introducing residuals, and it is easier to fit the residuals. However, from the perspective of backdoor attack, if the input x contains error information, the shortcut connection will save such error and pass it to the deeper network layer, resulting in the output of the network layer being misleading as well afterwards. Based on this characteristic, we design the BadRes block, as shown in Figure ~\ref{fig:block}b. The target output $H^*(x)$ is: 

\begin{equation}
\label{Equation:BadRes}
H^*(x) = F^*(x) - \alpha x
\end{equation}

This makes $F(x)$, which originally fit 0, become $F^*(x) = H^*(x) + \alpha x$, which also makes the trigger features in the poisoned sample x easier to be learned by $F^*(x)$. Then the output is given to the benign residual network afterwards, and as mentioned before, the deep residual connections provide a shortcut for backdoor attacks since the $H^*(x)$ containing the error information is always retained. This approach enables the model to learn the trigger features more quickly without changing the feature distribution of the benign inputs, and thus is not easily detected by existing backdoor detection methods, as analyzed in the Experiments section.

\subsubsection{Poisoning data phase.}

In this phase, we follow the setup with other backdoor attack methods, where the attacker adds the trigger to part of the benign training dataset and modifies the corresponding label as the target label. Let $D$ represent all training data, and a certain percentage of them are poisoned samples, such a percentage is called poisoning rate $\gamma$. i.e., $\gamma = \frac{D_p}{D}$, where $D_p$ is the set of poisoned samples with the trigger added. For the poisoned sample $x'$, the target class is usually set to $y_t$, which is generated as shown in Equation ~\ref{Equation:trigger}

\begin{equation}
\label{Equation:trigger}
    x' = (1-\lambda)x + \lambda t
\end{equation}

The $t$ in the equation represents the trigger. Due to this way of pasting the trigger on the original input, the residual connection will preserve the trigger features, while the BadRes block will take advantage of the shortcut connection to amplify the trigger features.

\subsubsection{Training phase.}

\begin{algorithm}
\caption{Training process of BadRes}

\label{alg:train.BadRes}
\begin{algorithmic}

\STATE {\textbf{Data}: all labeled training data $\mathcal{X}$, and label $\mathcal{Y}$ }

\STATE {\textbf{Input}: embedding layers $P$; encoder layers $L$; the classifier $F$; poisoning layer index $k$; BadRes rate $\alpha$}

\WHILE{not done}
    \FOR{$(X_i, Y_i)$ \emph{in} $\mathcal{(X, Y)}$}
        \STATE $E_{X_i} \leftarrow P(X_i)$
        
        \FOR{\emph{encoder layer} $g_i$ \emph{in} $L$}
            \STATE $O_i \leftarrow g_i(X_i)$
            \IF{$i$ == $k$} 
                \STATE $E_{X_i} = O_i - \alpha E_{X_i}$
            \ELSE
                \STATE$E_{X_i} = O_i + E_{X_i}$
            \ENDIF
        \ENDFOR
        \STATE $\widehat{Y_i} \leftarrow F(E_{X_i})$
        \STATE $\ell_i = \text{CrossEntropy}(\widehat{Y_i}, Y_i)$
    \ENDFOR
    \STATE update $P, L, F$ with loss $\mathcal{L} = \sum_i{\ell_i}$
\ENDWHILE

\end{algorithmic}
\end{algorithm}

The attacker uses the standard training approach and the overall training process is shown in algorithm ~\ref{alg:train.BadRes}. During the encoding process, one layer of the residual connection is selected to be changed to BadRes block, and the rest of the network is not modified. For our poisoned model parameter $\theta '$, the objective is to solve the following optimization problem:

\begin{equation}
\label{Equation:ResNe_objective}
    \theta ' = \mathop{\arg\min}\limits_{\theta}E_{x\sim X,x'\sim X'}[\mathcal{L}(x,y,\theta)+\mathcal{L}(x',y',\theta ')]
\end{equation}

Where x and y are benign input and benign target output, respectively, $x'$ is the poisoned sample, $y'$ is the target label, and $\mathcal{L}$ means the loss function. The backdoor attack of BadRes is completed when the training is finished. In the inference phase, the model suffering from BadRes attack would perform normally on the benign test sample, and the output would be the target label after adding the backdoor trigger.

\section{Experimental Results and Discussion}
We mainly conduct experiments under the CV task because the residual connection is mainly applied in CV models, especially the recent mainstream models such as ViT~\cite{ViT2021} and BEiT~\cite{BEiT2021} based on the self-attentive mechanism. We will present the experimental details in this section.

\subsection{Experimental settings}
\label{sec: setting}
\subsubsection{Datasets.}
We conducted experiments on four datasets, including MNIST~\cite{lecun1998gradient}, CIFAR-100~\cite{Krizhevsky09learningmultiple}, Food-101~\cite{food101_2014}, and ImageNet~\cite{ILSVRC15}. These datasets cover simple datasets with fewer categories and complex datasets with more categories to ensure that the effectiveness of the method can be tested comprehensively. MNIST is a 10-classification dataset with black and white images of $28 \times 28$ pixel size, containing 49K training data and 1K test data. CIFAR-100 has 100 classes, each of which is a color image of size $32 \times 32$, including 50K training data and 10K test data. Food-101 has 101 classes with 1K images per category, of which 75,750 images are used for training and the rest for testing. For the ImageNet dataset, to simplify the testing, we randomly selected 10K of the data for training and 1K for testing.

\subsubsection{Victim Models.}
We choose three self-attention models that work well in CV tasks and use residual connections for our experiments. They are ViT~\cite{ViT2021}, the most classical of the visual Transformer models, DeiT~\cite{DeiT2021}, which adds a distillation mechanism to the former, and BEiT~\cite{BEiT2021}, which adds a Mask mechanism to the pre-training process. For all victim models, the patch size is $16 \times 16$.

\subsubsection{Baseline Method.}
We compare BadRes with two backdoor attack methods. One is the classic backdoor attack method BadNets~\cite{BadNets2017} which is based on the data poisoning approach, and the other is the Blended attack~\cite{Blended2017} where the trigger is more stealthy and the trigger is blended with the original image. Meanwhile, we choose two ways to test the stealthiness of the three attack methods. The first one is fine-tuning~\cite{finetune2014} which weakens the backdoor attack to a certain extent, and the second one is STRIP~\cite{STRIP2019} method which determines whether it contains a backdoor based on the input.

\subsubsection{Hyper-parameters and Training Details.}

\begin{figure}
\centering
\includegraphics[width=\linewidth]{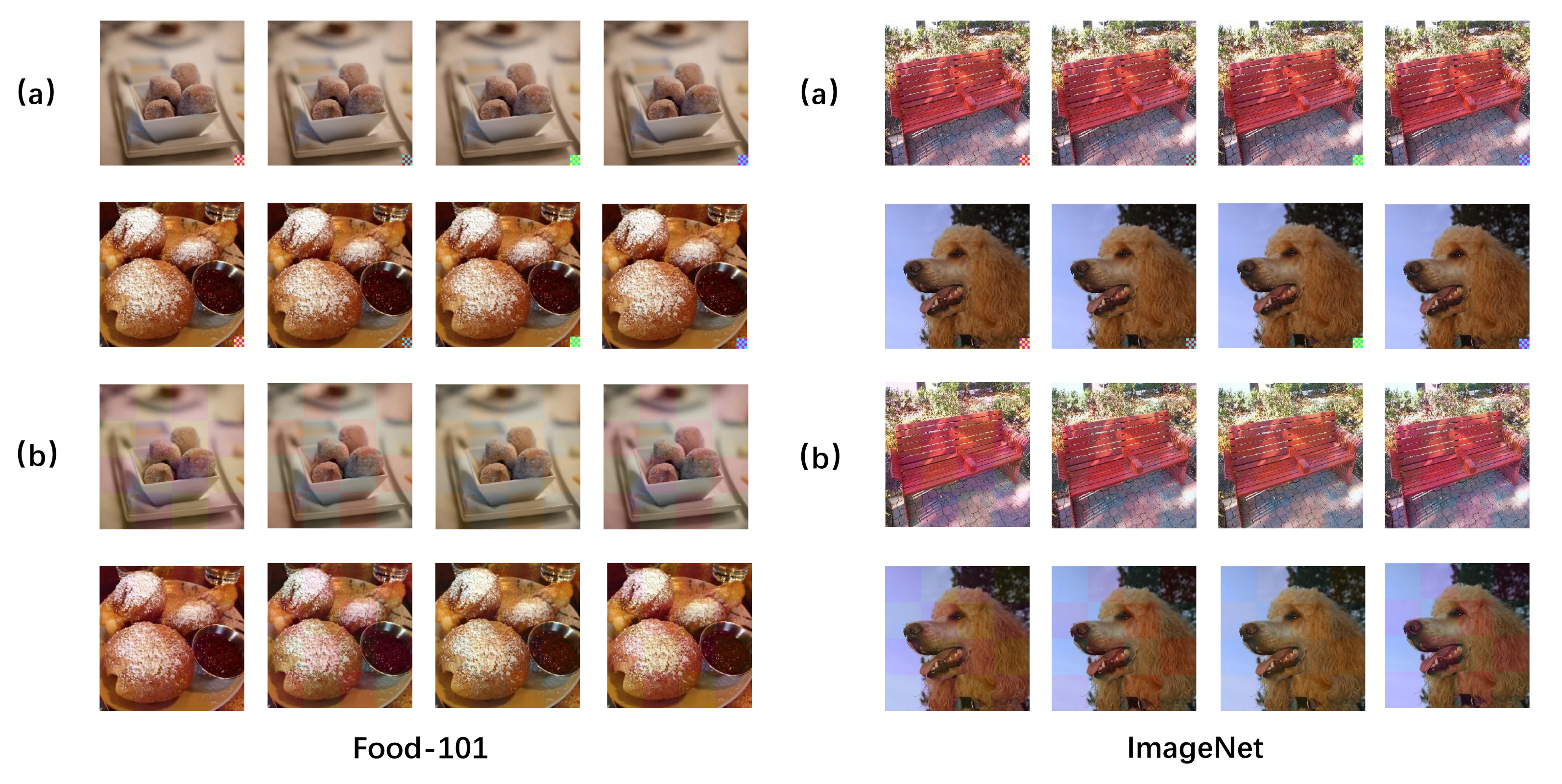}
\caption{Two different ways of generating poisoned data. A total of four different two-color grid-like shapes were tested as triggers. (a) for the BadNets and BadRes poisoned samples, the trigger is pasted in the lower right corner of the image. (b) is the poisoned sample of Blended attack, and the trigger is blended with the whole graph.}
\label{fig:trigger}
\end{figure}

For the experimental setup of the backdoor attack, we choose a poisoning rate of 10\% for all the backdoor attack datasets and the target labels are all 0. The poisoning data are shown in Fig.~\ref{fig:trigger}, all images are resized to $224 \times 224$. A total of four different bicolor grid maps are tested as triggers. In our BadRes method, the hyperparameter $\alpha$ in the BadRes block is 0.5. For the BadNets and BadRes methods, the poisoning samples are generated by pasting a trigger of size $16 \times 16$ in the lower right corner of the image. For the Blended method, the poisoned samples are generated by blending the triggers with the whole image, where the transparency of the triggers is 5\%. For the training process of all backdoor attacks, we set the initial learning rate to 0.0001, the batch size to 32, the epoch size to 30, and use the Adam optimizer~\cite{Adam2015}.

\subsubsection{Evaluation Metrics.}
We use two metrics that are consistent with BadNets~\cite{BadNets2017}. We evaluate the effect of the attack in two aspects. For the case where the input does not add the trigger, we use the benign accuracy (BA) to evaluate the different attacks. Where the higher BA indicates that the method is better hidden and does not affect the model performance in normal use. For the case of adding triggers to the inputs, we use the attack success rate (ASR) to evaluate the attacks, as shown in Equation~\ref{Equation:ASR}.

\begin{align}
\label{Equation:ASR}
    ASR = \frac{\#\mathrm{(misclassified~samples)}}{\#\mathrm{(total~poisoned~samples)}}
\end{align}
which is the rate of successfully misleading the model output to the target class among all the poisoned samples.

\subsection{Attack effectiveness}

We follow the experimental setup in the previous section to experiment. In order to fairly and comprehensively evaluate the effectiveness of BadRes, we test it separately for different datasets with different victimization models.

\setlength{\tabcolsep}{4pt}
\begin{table}
\begin{center}
\caption{The comparison of different methods against Vision Transformer without defense on the MNIST, CIFAR100, Food-101 and ImageNet dataset. Among all attacks, the best result is denoted in boldface.}
\label{table:different_datasets}
\begin{tabular}{l|ll|ll|ll}
\hline\noalign{\smallskip}
 Method & \multicolumn{2}{c|}{BadRes} & \multicolumn{2}{c|}{BadNets} & \multicolumn{2}{c}{Blended}\\
 Datasets & ASR  & BA  & ASR  & BA  & ASR  & BA \\
\noalign{\smallskip}
\hline
\noalign{\smallskip}
MNIST  & 97.67 & 99.42 & \textbf{99.74} & 99.26 & 98.25 & \textbf{99.52}\\
CIFAR-100 & \textbf{95.96} & \textbf{92.37} & 95.57 & 91.99 & 95.42 & 90.72\\
Food-101 & \textbf{98.57} & \textbf{83.43} & 86.8 & 79.6 & 96.52 & 82.16\\
ImageNet & \textbf{100} & \textbf{90.67} & 84.28 & 87.8 & 98.24 & 87.42\\
\noalign{\smallskip}
\hline
\noalign{\smallskip}
Average & \textbf{98.05} & \textbf{91.47} & 91.60 & 89.66 & 97.11 & 89.91\\
\noalign{\smallskip}
\hline
\end{tabular}
\end{center}
\end{table}
\setlength{\tabcolsep}{1.4pt}
Firstly, the detailed results of the experiments for different datasets are shown in Table ~\ref{table:different_datasets}. We use three backdoor attacks for the ViT model, and the experimental results are the average of four different triggers shown in Figure ~\ref{fig:trigger}. For the datasets with low image pixels such as MNIST and CIFAR100, the ASR of the three backdoor attack methods are similar and have less impact on the results of clean data. As the image pixels increase, the ASR of the BadNets method gradually starts to decrease, dropping to about 85\% on the Food101 and ImageNet datasets. Also, the BadNets method has an impact on BA, especially on the Food101 dataset, where BA drops to 79\%. We analyze that this is because, at the higher pixel level, this pasted trigger is in a small percentage and may only affect part of the patch embeddings for ViT. This makes the model unable to learn the features of the target label during the forward propagation, and the original correct feature extraction is also affected. In contrast, BadRes and Blended are more stable than BadNets, and BadRes is better in overall average ASR and BA. ASR can reach almost 100\%, which also indicates that our designed BadRes block can amplify the error message in patch embeddings. On the ImageNet dataset, the BA of BadRes can be 2 percentage points higher than Blended, which can also show that our proposed method has almost no effect on the classification results of benign data.

Secondly, we experiment on the imageNet dataset for three different models, and the results are shown in Table ~\ref{table:different_models}. For the ViT model, the results are consistent with the above experimental results on different datasets, and the BadRes and Blended methods are more effective than BadNets. For the BEiT model, we find that Blended works best and the ASR reaches 90.75\%, while BadNets is less effective. We analyze that this is because BEiT uses the mask mechanism for pre-training, which can defend against pasted backdoor attacks to a certain extent. The Blended backdoor attack, on the other hand, is less affected because the trigger covers the whole image. And our BadRes method is still able to reach 90\% ASR with the same triggers as BadNets, which also shows that the BadRes method successfully strengthens the backdoor injection of error messages. For DeiT using the distillation mechanism, the attack effectiveness of both BadNets and Blended decreases to a certain extent, while the BadRes' ASR can still reach 100\%. This indicates that our proposed method is hardly affected by distillation. Overall, for the three different victimization models, our proposed method is more threatening and has less impact on the performance of the original model.

\setlength{\tabcolsep}{4pt}
\begin{table}
\begin{center}
\caption{The comparison of different victim models without defense on the ImageNet dataset. Among all attacks, the best result is denoted in boldface.
}
\label{table:different_models}
\begin{tabular}{l|ll|ll|ll}
\hline\noalign{\smallskip}
 Method & \multicolumn{2}{c|}{BadRes} & \multicolumn{2}{c|}{BadNets} & \multicolumn{2}{c}{Blended}\\
 Model & ASR  & BA  & ASR  & BA  & ASR  & BA \\
\noalign{\smallskip}
\hline
\noalign{\smallskip}
ViT  & \textbf{100} & \textbf{90.67} & 84.28 & 87.8 & 98.24 & 87.24\\
BEiT & 90.67 & 83 & 73.86 & \textbf{87.88} & \textbf{93.75} & 85.36\\
DeiT & \textbf{100} & \textbf{83.33} & 75.9 & 82.14 & 94.35 & 83.24\\
\noalign{\smallskip}
\hline
\noalign{\smallskip}
Average & \textbf{96.89} & 85.67 & 78.01 & \textbf{85.94} & 95.45 & 85.28\\
\noalign{\smallskip}
\hline
\end{tabular}
\end{center}
\end{table}
\setlength{\tabcolsep}{1.4pt}

\subsection{Attack stealthiness}

\begin{figure}[h]
  \centering
  \includegraphics[width=\linewidth]{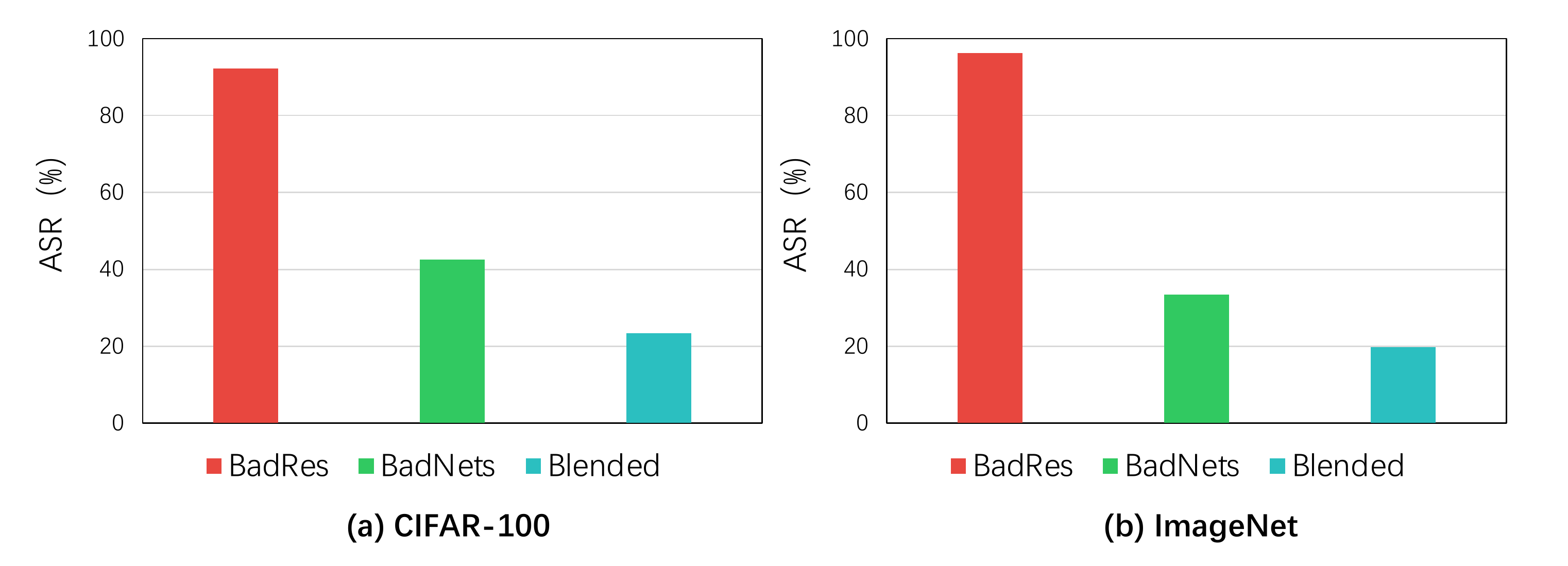}
  \caption{Backdoor attacks are eliminated by fine-tune on the benign dataset. The table shows the ASR after fine-tune, and higher values mean that the method is less affected by fine-tune.}
  \label{fig:fine-tune}
\end{figure}

Firstly, we choose fine-tune to test whether BadRes is easily eliminated. Since fine-tune is performed on benign datasets, the backdoor-related neuron parameters are easily lost during the learning process, and the results are shown in Fig.~\ref{fig:fine-tune}. BadRes maintains its ASR above 90\% after fine-tuning on the CIFAR-100 and ImageNet datasets. On the other hand, the ASR of BadNets drops to around 40\%. And Blended is affected the most, with ASR dropping to 23.4\% and 19.81\% on the two datasets, respectively. This indicates that BadRes is more difficult to be eliminated.

\begin{figure}
  \centering
  \includegraphics[width=\linewidth]{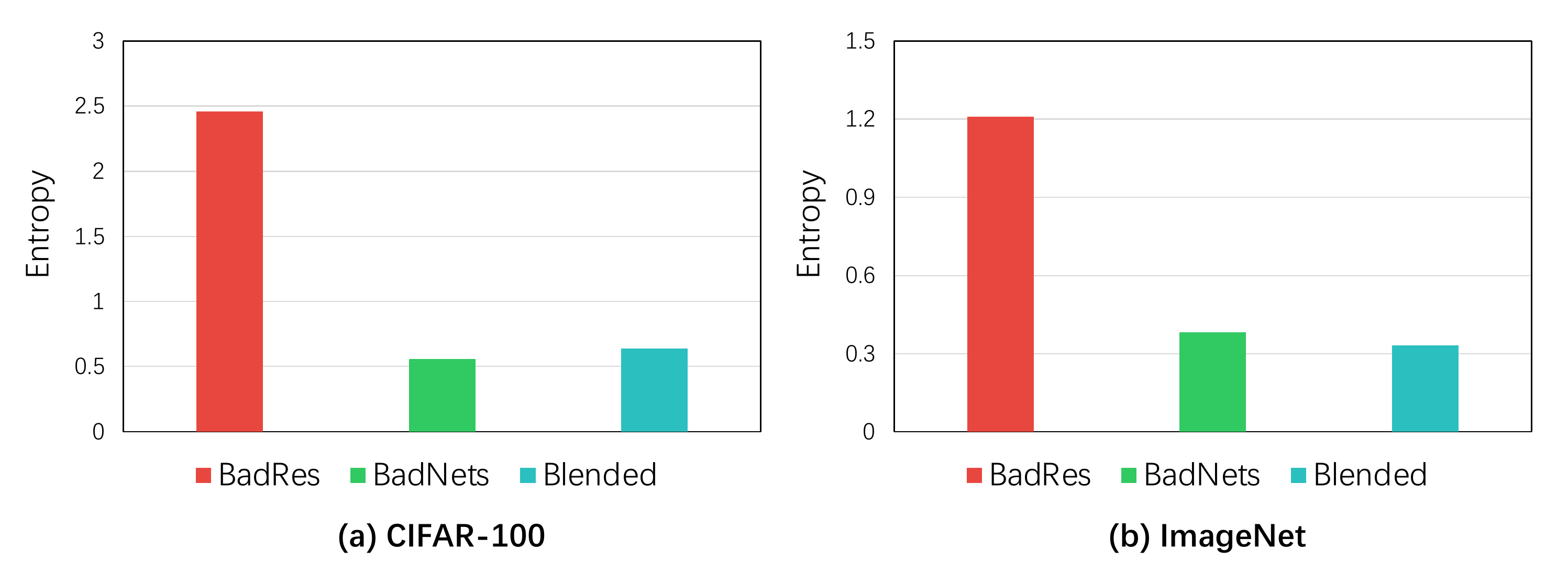}
  \caption{Entropy of different backdoor attack methods generated based on STRIP~\cite{STRIP2019} method. higher Entropy means harder to be detected that the model suffers from backdoor attack.}
  \label{fig:strip}
\end{figure}

Secondly, we choose the STRIP method to test the stealthiness of different backdoor attack methods, which detects the backdoor attack by adding perturbations to the input and then calculating the average entropy of the predictions, where the lower entropy indicates that the model is more likely to have been injected with a backdoor. We choose the ViT model to calculate the entropy after different attacks on two datasets, MNIST and CIFAR-100, and the results are shown in Fig.~\ref{fig:strip}. The entropy of our proposed method on both datasets is much higher than that of BadNets and Blended, which also indicates that BadRes is more difficult to be detected by this defense method and has better stealthiness.

\subsection{Ablation studies}

\subsubsection{Hyperparameter and comparison structure experiments of BadRes Block.}

\setlength{\tabcolsep}{4pt}
\begin{table}[h]
\begin{center}
\caption{The impact of hyperparameters $\alpha$ in BadRes Block, and hyperparameters $\beta$ of comparison structure experiments on ASR and BA. Experiments are chosen on two datasets, ImageNet and Food101.}
\label{table:different hyperparameters}
\begin{tabular}{ll|llll|llll}
\hline\noalign{\smallskip}
    \multirow{2}{*}{Dataset} & \multirow{2}{*}{Metircs} & \multicolumn{4}{c|}{Weight $\alpha$} & \multicolumn{4}{c}{Weight $\beta$} \\
 & & 0.2 & 0.4 & 0.6 & 0.8 & 0.2 & 0.4 & 0.6 & 0.8\\
\noalign{\smallskip}
\hline
\noalign{\smallskip}
		
\multirow{2}{*}{ImageNet} & ASR & 100 & 100 & 100 & 99.75 & 75.03 & 77.54 & 78.82 & 80.93 \\
& BA & 85.16 & 87.44 & 90.83 & 89.42 & 33.84 & 26.35 & 20.81 & 17.48 \\
\noalign{\smallskip}
\hline
\noalign{\smallskip}
\multirow{2}{*}{Food-101} & ASR & 98.71 & 98.45 & 98.19 & 98.01 & 67.41 & 68.49 & 70.59 & 73.22 \\
& BA & 81.28 & 82.43 & 83.57 & 83.12 & 27.88 & 22.43 & 19.74 & 17.06\\
\noalign{\smallskip}
\hline
\end{tabular}
\end{center}
\end{table}
\setlength{\tabcolsep}{1.4pt}

In this subsection, we discuss the effect of the alpha parameter in Equation~\ref{Equation:BadRes} on the results. Also, in order to reveal the reason why BadRes can be successfully attacked, we design a comparison structure with the objective function shown in Equation~\ref{Equation:comparison}:

\begin{equation}
\label{Equation:comparison}
H'(x) = x - \beta F'(x)
\end{equation}

Unlike the objective function of the BadRes Block, the output of the comparison structure completely retains the output $x$ of the previous layer and subtracts the result $F'(x)$ computed in this layer. The purpose of this setup is to form a comparison experiment with the BadRes block to explore how $x$ and $f(x)$ are affected by the backdoor information. We experiment on two datasets, ImageNet and Food101, with more categories and higher pixels. For$\alpha$ and $\beta$, we choose 5 representative values between 0 and 1, and the other parameters are set the same as in the previous section. The results are shown in the Table~\ref{table:different hyperparameters}.

For different hyperparameters $\alpha$, the ASR is stable around 100\% and 98\% on the two datasets. This indicates that the backdoor features can be learned by $f^*(x)$ and partial information of $x$. However, it is unable to determine exactly which part is more important, and we will analyze it in the next part, combining the experimental results of the comparison structures. On the other hand, BA improves slightly with increasing alpha. We analyze that this is because $f^*(x)$ needs to learn the benign information in the previous output $\alpha x$ in addition to $H^*(x)$, and thus can improve BA. However, this would increase the cross entropy between the trigger and benign input, which would sacrifice a certain stealthiness of BadRes. Therefore this parameter setting needs to be based on the specific needs of the attacker.

In the experiments comparing the structures, the output results are significantly influenced by benign inputs. When beta is 0.8, BA is below 20\% on both datasets. This is due to the fact that $f'(x)$ becomes $\frac{(x - H'(x))}{\beta}$, which is opposite to the original $H(x) - x$, resulting in a decrease in the original correct performance. For the poisoned data, the results show that this objective function enables the model to learn some of the backdoor features, and is able to achieve the ASR of 75\% on the ImageNet dataset and 65\% on the Food101 dataset, with a small increase in ASR as the $\beta$ increases. This indicates that the backdoor features are retained in the output $x$ of the previous layers all the time and are still able to deliver such error messages even when the $\beta$ is small, which warns us that a jump structure like residual connection can be its weakness. Also, the fact that BadRes can achieve the ASR close to 100\% compared to the comparison structure indicates that our proposed design is indeed able to amplify the backdoor error message.

\subsubsection{The selection of poisoned layer index.}

\begin{figure}[h]
  \centering
  \includegraphics[width=\linewidth]{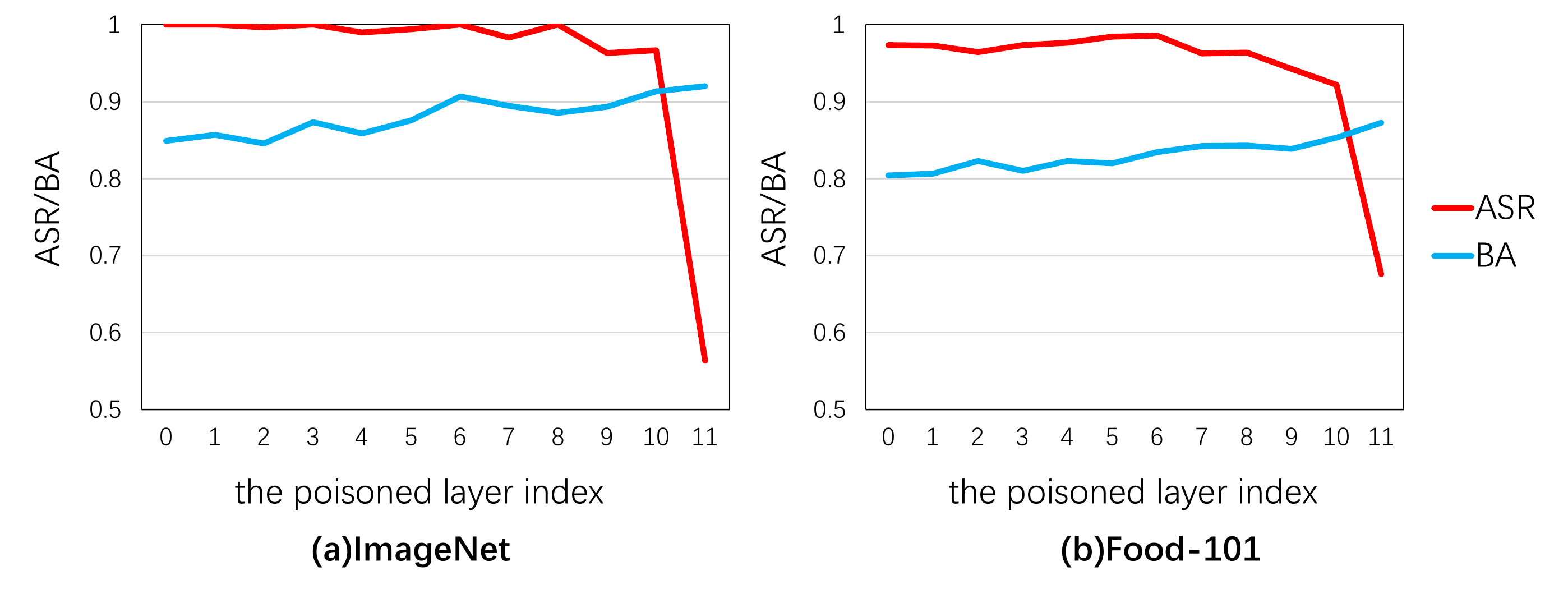}
  \caption{Impact of BadRes' poisoning layer index selection on ASR and BA. Figure (a) shows the results on the ImageNet dataset and Figure (b) shows the results on the Food-101 dataset. The red line is the ASR and the blue line is the BA. Our results show that the poisoned layer index decreases sharply at the last two layers for ASR and is stable for the rest of the choices. There is a small increase in BA as the poisoned layer index increases.}
  \label{fig:layer}
\end{figure}

In this section, we discuss the impact of BadRes' poisoned layer index on ASR and BA. We conduct experiments on both ImageNet and Food-101 datasets, and the hyperparameters are consistent with Section~\ref{sec: setting} except that the poisoned layer index is changed. The results are shown in Fig.~\ref{fig:layer}. When the layer index is less than or equal to 9, the ASR is relatively stable and close to 100\%. When the layer index is the last two layers, the ASR drops sharply, especially when the last layer is selected, the ASR drops to below 60\% in the ImageNet dataset and below 70\% in the Food-101 dataset. This phenomenon indicates that when the layer index is larger, the clean residual network layers have already extracted the correct features, and the last poisoned layer has difficulty in producing a large bias on the feature vector. This also demonstrates that when the index of the poisoned residual unit is small, it can pass down the wrong information and produce a domino effect, thus achieving a higher ASR. On the other hand, the BA rises as the poisoning layer index increases. This also indicates that the larger the poisoning layer index in the case of clean data, the less the impact of the model's ability to extract the correct features.

\subsubsection{Effect of poisoning rate on the ASR and BA of BadRes.}

\begin{figure}
  \centering
  \includegraphics[width=9cm]{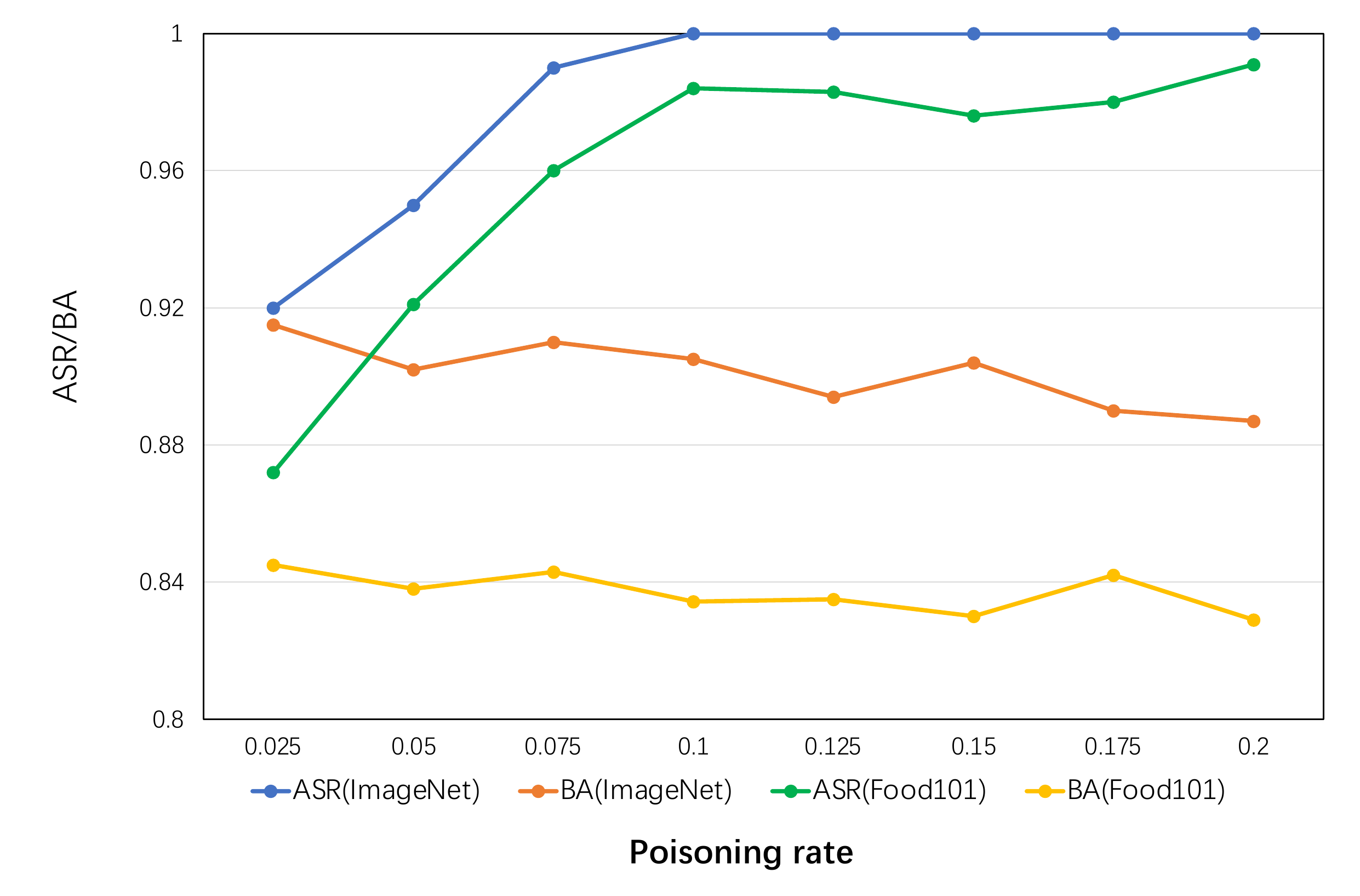}
  \caption{Effect of the selection of poisoning rate on ASR and BA. We choose the poisoning rate in the range of 0.025 to 0.2 for our experiments on both ImageNet and Food101 datasets. In this figure, the parameters are kept constant except for the variation of the virulence rate. We observe that the ASR stabilizes after the poisoning rate reaches 0.1, while the BA is almost unaffected by the change in the poisoning rate.}
  \label{fig:poisoning_rate}
\end{figure}

To discuss the effect of the poisoning rate on BadRes, consistent with the previous section, we conduct experiments on both datasets and varied only the poisoning rate. As shown in the Fig.~\ref{fig:poisoning_rate}, the ASR exceeds 90\% when the poisoning rate reaches 0.05. This result shows that BadRes can learn the features of triggers with only a small amount of poisoning data. On the other hand, the BA dose not decrease much with the increase of poisoning rate, stabilizing at around 90\% and 84\% on the ImageNet and food101 datasets. This also indicates that our proposed method has good stability.

\subsubsection{Impact of target label setting.}

\setlength{\tabcolsep}{4pt}
\begin{table}[h]
\begin{center}
\caption{The impact of target label setting on our attack methods. We select five target labels, 0, 1, 2, 4 and 8, and test them on four datasets. We observe that different target labels in the same dataset have little effect on the ASR and BA of our method.}
\label{table:different target}
\begin{tabular}{l|ll|ll|ll|ll}
\hline\noalign{\smallskip}
 Dataset & \multicolumn{2}{c|}{MNIST} & \multicolumn{2}{c|}{CIFAR-100} & \multicolumn{2}{c|}{Food-101} & \multicolumn{2}{c}{ImageNte}\\
 Label & ASR  & BA  & ASR  & BA  & ASR  & BA & ASR & BA\\
\noalign{\smallskip}
\hline
\noalign{\smallskip}
		
label=0  & 97.67 & 99.42 & 95.96 & 92.37 & 98.57 & 83.43 & 100 & 90.67\\
label=1 & 97.78 & 99.32 & 95.74 & 92.59 & 98.45 & 83.51 & 99.56 & 90.75\\
label=2 & 96.86 & 99.54 & 96.04 & 92.11 & 98.14 & 83.79 & 100 & 90.34\\
label=4 & 98.41 & 99.36 & 96.59 & 91.87 & 98.78 & 83.11 & 99.78 & 90.76\\
label=8 & 97.04& 99.28 & 95.31 & 92.17 & 98.09 & 83.87 & 99.87 & 90.89\\
\noalign{\smallskip}
\hline
\end{tabular}
\end{center}
\end{table}
\setlength{\tabcolsep}{1.4pt}

In this experiment, we discuss the effect of target label setting on BadRes performance. The results are detailed in Table~\ref{table:different target}, where we test setting five different target labels for the poisoned data in four datasets. The results show that the ASR and BA of different target labels are similar in the same dataset. This also indicates that BadRes' performance is largely unaffected by the target label settings.

\section{Conclusion and future works}

In this paper, we propose BadRes, a backdoor attack method against residual connections. since the shortcut connection method passes the trigger features, we design a method that amplifies the trigger features, and this attack method is extremely threatening to models that use residual connections. We test the effectiveness of BadRes on four popular datasets and three mainstream victim models, and it outperforms BadNets and Blended in terms of average ASR and BA. Meanwhile, we use fine-tuning and STRIP to test the stealthiness of our method, and the results demonstrate that BadRes is more difficult to be detected and eliminated compared to the other two baseline models. We then use ablation and comparison experiments to reveal the reason for the success of BadRes and the reason why residual connections are so weak.

Finally, we hope that our work will bring attention to the security of residual connections and that practitioners should be aware of the potential risks when applying this technique. On the other hand, how to improve the residual connection structure to resist such a backdoor attack deserves further research.

%
%
%
\bibliographystyle{splncs04}
\bibliography{main}
%




\end{document}